\documentclass[a4paper,fleqn]{cas-sc}

\usepackage[numbers]{natbib}
\usepackage{makecell}
\usepackage{xcolor, colortbl}
\usepackage{array}
\usepackage{appendix}

\usepackage{longtable}
\AtBeginEnvironment{longtable}{\sffamily}

\definecolor{Gray}{gray}{0.9}
\newcommand{\ie}{\emph{i.e.},~}
\newcommand{\eg}{\emph{e.g.},~}
\newcommand{\ea}{\emph{et al.}~}

\def\tsc#1{\csdef{#1}{\textsc{\lowercase{#1}}\xspace}}
\tsc{WGM}
\tsc{QE}

\begin{document}
\let\WriteBookmarks\relax
\def\floatpagepagefraction{1}
\def\textpagefraction{.001}

\shorttitle{Towards better Interpretable and Generalizable AD detection using Collective Artificial Intelligence}    

\shortauthors{HD Nguyen}  

\title [mode = title]{Towards better Interpretable and Generalizable AD detection using Collective Artificial Intelligence}

%

\author[1]{Huy-Dung Nguyen}[orcid=0000-0002-3980-8029]
\cormark[1]

\ead{huy-dung.nguyen@u-bordeaux.fr}

\author[1]{Michaël Clément}

\author[1]{Boris Mansencal}

\author[1]{Pierrick Coupé}

\affiliation[1]{organization={Univ. Bordeaux, CNRS, Bordeaux INP, LaBRI, UMR 5800},
    postcode={33400 Talence},
    country={France}}

\cortext[1]{Corresponding author}


\begin{abstract}
Alzheimer’s Disease is the most common cause of dementia. Accurate diagnosis and prognosis of this disease are essential to design an appropriate treatment plan, increasing the life expectancy of the patient. Intense research has been conducted on the use of machine learning to identify Alzheimer's Disease from neuroimaging data, such as structural magnetic resonance imaging. In recent years, advances of deep learning in computer vision suggest a new research direction for this problem. Current deep learning-based approaches in this field, however, have a number of drawbacks, including the interpretability of model decisions, a lack of generalizability information and a lower performance compared to traditional machine learning techniques. In this paper, we design a two-stage framework to overcome these limitations. In the first stage, an ensemble of 125 U-Nets is used to grade the input image, producing a 3D map that reflects the disease severity at voxel-level. This map can help to localize abnormal brain areas caused by the disease. In the second stage, we model a graph per individual using the generated grading map and other information about the subject. We propose to use a graph convolutional neural network classifier for the final classification. As a result, our framework demonstrates comparative performance to the state-of-the-art methods in different datasets for both diagnosis and prognosis. We also demonstrate that the use of a large ensemble of U-Nets offers a better generalization capacity for our framework.
\end{abstract}




\maketitle

\section{Introduction}

\subsection{Context}
Alzheimer's Disease (AD) is a common neurodegenerative disease characterized by the progressive impairment of cognitive functions. This pathology is the most common type of dementia and a major cause of mortality in people over 65 years old \cite{wong_economic_2020}. Memory loss is the first symptom of AD, and it gets worse over time. As the disease progresses, AD patients require help even with basic activities, making a significant impact on their daily lives as well as their family. In 2006, there were 26.6 million AD patients worldwide~\cite{brookmeyer_forecasting_2007} which increased to 46.8 million in 2015. Moreover, this number is expected to reach 131.5 million in 2050 \cite{herrera_world_2016}. As a result, the costs of caring for Alzheimer's patients are rapidly increasing. Furthermore, more treatments and services are required over time, continuously driving up those costs. Consequently, early and accurate detection of Alzheimer's Disease is critical for the development of new therapies, slowing disease progression, and reducing associated costs.

The prodromal stage of AD is Mild Cognitive Impairment (MCI) \cite{markesbery_neuropathologic_2010}. People may experience minor changes in cognitive abilities at this stage but there is still no impact on their daily lives \cite{albert_diagnosis_2011}. Statistically, 10 – 17\% of people with MCI will progress to AD over a few years while other MCI patients will remain stable \cite{hamel_trajectory_2015}. The first group refers to progressive MCI (pMCI) and the second one refers to stable MCI (sMCI). Besides the need of distinguishing AD patients from cognitively normal people (CN) (\ie AD diagnosis), identifying pMCI patients from sMCI patients (\ie AD prognosis) is even more crucial to apply appropriate therapies and slow down the transition from MCI to AD. Therefore, a fast and accurate tool for both AD diagnosis and prognosis is expected to help clinician to take care of the patient as soon as possible.

Brain atrophy is an important biomarker of Alzheimer’s disease. Many studies state that this morphological change may occur before the first cognitive symptoms of AD \cite{gordon_spatial_2018, jung_deep_2021, coupe_detection_2015, bron_cross-cohort_2021}. Those anatomical changes can be identified with the help of structural magnetic resonance imaging (sMRI) \cite{bron_standardized_2015}. Recently, with the advances of convolutional neural networks (CNN), a large number of methods have been proposed for automatic AD diagnosis and prognosis using sMRI \cite{wen_convolutional_2020, al-shoukry_alzheimers_2020, ebrahimighahnavieh_deep_2020}. Indeed, over the last decade, Deep Learning (DL) has demonstrated breakthrough performance in natural image classification. Unlike traditional machine learning algorithms, deep learning allows to automatically extract discriminative features from the input without prior knowledge. It has also received a lot of attention in medical imaging analysis, where it can help clinicians to follow disease progression. However, the large size of 3D sMRI and the limited GPU memory has required to adapt DL methods to medical imaging. These methods can be categorized into: 2D slice-based methods, 3D subject-based methods, 3D region-of-interest (ROI) methods and 3D patch-based methods.

\subsection{Related works}

\textbf{2D slice-based methods:} The concept behind slice-based approaches is that a minimal number of 2D slices may accurately depict the disease status. Some methods employ their own strategy to extract the most appropriate 2D slices from 3D sMRI, while others use standard image projections (\ie coronal, sagittal, axial plane) \cite{ebrahimighahnavieh_deep_2020}. Valliani \ea considered only the median axial slice of sMRI and used ResNet for AD diagnosis \cite{valliani_deep_2017}. Pan \ea trained 123 classifiers for 123 2D slice positions of three projection planes for both AD diagnosis and prognosis \cite{pan_early_2020}. The 15 models with the highest accuracy on validation set were chosen to form the final ensemble model. Qiu \ea manually chose three slice positions to analyze well-known regions associated with Alzheimer's disease: lateral ventricles, inferior temporal, and middle temporal cortices \cite{qiu_fusion_2018}. Three CNN models (one per region) were then trained for the problem of classification CN vs. MCI. The final result was based on the majority vote. Entropy-based sorting is another method to select slice positions. It is based on the hypothesis that a slice with higher intensity variation is more informative. This strategy was used in \cite{hon_towards_2017, jain_convolutional_2019} to select the most 32 informative slices for various AD classification tasks. All of these slice-based methods have the advantage of being based on well-known CNNs architectures dedicated to natural image classification. However, a comparative study showed that 2D slice-based methods were less efficient than 3D methods \cite{wen_convolutional_2020}. This study explained that spatial information is not fully exploited by 2D slice-based methods which limits their performance.

\textbf{3D subject-based methods:} Recently, more methods using the whole 3D MRI have been proposed for AD classification (3D subject-based methods). In general, these models have fewer layers than 2D slice-based approaches due to the limited computing capacity. Backström \ea used a 3D CNN with 8 layers for AD diagnosis \cite{backstrom_efficient_2018}. Yee \ea used dilated convolution to increase the model depth to 11 layers \cite{yee_construction_2021}. In doing so,  they improved the receptive field while keeping a reasonable number of parameters. VGG and ResNet are usually employed by many authors for classification tasks in natural images. In \cite{korolev_residual_2017}, the authors implemented 3D version of these two architectures and showed comparable performance of both models for different AD classification tasks. Modern architecture like inception module was also proposed in \cite{oh_classification_2019}. Li \ea proposed a multi-model for AD diagnosis \cite{li_alzheimers_2017}. As each model had a different receptive field, the ensemble model was expected to be able to capture both global and local features. Overall, 3D subject-based methods have the advantage of preserving spatial information. However, since the 3D architectures are shallower, with current memory limitations these models do not yet offer optimal performance.

\textbf{3D regions-of-interest (ROI) methods:} With a limited computing capacity, reducing the input dimension is a good way to increase the model complexity. Many methods focused on particular parts of the brain known to be related to AD. Only one or a few small 3D cubic sub-volumes located at specific brain structures are used as input. Consequently, deeper models can be used and more complex patterns can be captured. The hippocampal region is a ROI well-known to be affected by AD \cite{schuff_mri_2009}. Huang \ea cropped a region centered at the hippocampi from sMRI. They used a VGG-like architecture for classification \cite{huang_diagnosis_2019}. Cui \ea used two cubic sub-volumes surrounding the left and right hippocampus to exploit also their adjacent regions for accurate AD classification \cite{cui_hippocampus_2019}. They suggested that these areas, including the parahippocampus and amygdala, may be involved in AD. The main drawback of this type of method is that they only use the information around a priori defined anatomical regions. In contrast, alterations caused by AD can affect other brain areas \cite{wachinger_whole-brain_2016}. Therefore, relevant information outside of the selected ROIs is not used, limiting model performance.

\textbf{3D patch-based methods:} Another way to reduce the input dimension is to use 3D patch-based methods. An MRI is simply divided into multiple smaller patches, all of them are then used for training. Cheng \ea extracted 27 overlapping patches that were uniformly distributed across the whole brain. They then trained 27 models (one per patch) and an ensemble model aggregating patch-level results to make the final decision \cite{cheng_classification_2017}. Li \ea divided the original MRI into 27 non-overlapping patches \cite{li_alzheimers_2018}. These patches were grouped into different clusters and one CNN was trained per cluster for the AD diagnosis problem. The final decision was made by ensembling these models. In several studies, Liu \ea used a landmark detection algorithm to locate the most informative patches in sMRI \cite{liu_anatomical_2018, liu_landmark-based_2018, liu_deep_2017}. In \cite{liu_anatomical_2018}, the authors trained 27 different models (one per patch) for the classification problem. The final decision was obtained by majority voting strategy. In \cite{liu_landmark-based_2018}, they designed an end-to-end CNN model with multiple branches, each one analyzing one patch. The learned features were concatenated and forwarded through a final CNN for AD classification. In \cite{liu_deep_2017}, they constructed multi-channels input from extracted patches and used a simple CNN for AD classification. Lian \ea performed a voxelwise anatomical correspondence across all available images \cite{lian_hierarchical_2020}. They then selected 120 voxel locations and used them as centers for extracting 120 patches. They built a single end-to-end CNN model in which feature representations learned from patch-level was concatenated at regional-level, feature representation at regional-level was then concatenated to provide the decision at subject-level. From a literature review, it appears that a single model is not enough to capture the diverse patterns of all patch locations \cite{wen_convolutional_2020}. Indeed, methods using multiple models \cite{cheng_classification_2017, li_alzheimers_2018, liu_anatomical_2018} offer better AD classification accuracy. Compared to previously detailed strategies, 3D patch-based methods enable to fully exploit the 3D information, to drastically reduce memory requirement and to analyze the entire MRI.

\subsection{Current limitations of DL in AD classification}
Although many efforts were made to adapt deep learning methods to AD classification, existing methods still present several limitations. Indeed, current approaches have limited prognosis performance and usually suffer from a lack of generalization and interpretability.

\textbf{Limited Performance:} At the time of writing this paper, CNN based-models seemed not to perform better than conventional machine learning methods (\eg support vector machine SVM). Bron \ea showed similar performance between CNN and SVM models while carefully following the state-of-the-art CNN designs \cite{bron_cross-cohort_2021}. In another study, Wen \ea \cite{wen_convolutional_2020} even found that their linear SVM model was at least as good as the best CNN model for AD diagnosis and better for AD prognosis. They both suggested that a more sophisticated DL architecture may help for better performance.

\textbf{Limited Generalization:} A recent survey showed that about 90\% of studies use the same dataset (\ie ADNI dataset) to evaluate their model performance which limits our knowledge of CNN performance on other databases \cite{ebrahimighahnavieh_deep_2020}. Moreover, most of the studies mentioned above used the same dataset for training and testing. Such validation framework is known to over-estimate method performance. Indeed, in-domain validation is dangerous as methods showing high performance on a single dataset might just better capture the particular characteristics of that dataset and might poorly perform in another dataset \cite{thibeau-sutre_mri_2022}. As a consequence, current DL literature offers limited knowledge about the generalization capability of DL methods on external datasets. This limitation does not only apply to AD classification application but also to other diseases (\eg Frontotemporal Dementia \cite{termine_reproducible_2022}, Parkinson's disease \cite{mei_machine_2021}, etc.). A general cause leading to a low generalization capacity is overfitting on the training set \cite{wen_convolutional_2020}. Especially, this often occurs when the size of the training domain is too small. To alleviate this problem, we applied several data augmentation techniques during the training process (see Section \ref{section:implementation_details}), making the model more robust to heterogeneity. Furthermore, our use of a large number of models (see Section \ref{section:collective_AI}) that can be seen as an ensemble model allows the reduction of generalization error~\cite{kotu_data_2015}.

\textbf{Limited Interpretation and Explanation:} Besides the need for an accurate and generalizable AD classification model, understanding the model decision is also vital. In this study, we consider two terminologies: interpretability and explainability as in \cite{barredo_arrieta_explainable_2020}. Interpretability refers to the passive characteristic of a model that can be directly understood by humans. By contrast, explainability refers to external procedures applied to a model to discover its internal functionalities. The majority of current deep learning methods use an external explainable method (\eg Class Activation Mapping, Gradient Class Activation Mapping, Guided Backpropagation) to study their model decision. However, some explainable methods (\ie Guided Backpropagation and Guided Gradient Class Activation Mapping) produce visually and quantitatively similar explanations between a model randomly-initialized and a trained model. This makes analysis based on the produced explanations suspicious \cite{adebayo_sanity_2020}. In \cite{yee_construction_2021}, two explainable methods were applied to the same model but different results were obtained. Moreover, in \cite{bron_cross-cohort_2021}, the obtained saliency map showed regions known to be little affected by AD. Indeed, each explainable method works differently, so the discovery may not be unique or little informative. For an interpretable model instead, humans can directly infer its characteristic without losing information due to additional actions. Thus, this kind of method seems to be more valuable to understand the model decision. However, to the best of our knowledge, there is currently few interpretable methods for AD classification, with our definition \cite{bass_NEURIPS2020, bass2021icam}.

\subsection{Contributions}
In this paper, to address these current major limitations of DL methods, we propose a novel interpretable, generalizable and accurate deep framework for both AD diagnosis and prognosis. This clinical tool is available at \url{https://volbrain.net}.

First, we propose a novel Deep Grading (DG) biomarker to improve the interpretability of deep model outputs. Inspired by the patch-based grading frameworks \cite{coupe_detection_2015, coupe_simultaneous_2012, tong_novel_2017, hett_multi-scale_2021, coupe_scoring_2012, hett_adaptive_2018}, this new biomarker can capture CN, AD patterns from MRI input and provides a grading map with a score between $-1$ and 1 at each voxel that reflects the disease severity. This interpretable biomarker may assist clinicians in localizing brain regions affected by AD, allowing them to make more informed decisions. 

Second, we propose to extend the concept of Collective Artificial Intelligence (AI) to AD diagnosis and prognosis. The collective AI consists of using a large number of communicating neural networks, each of them is specializing in a unique brain location. The global result is then obtained by fusing the local results. For the brain segmentation application, it has demonstrated a better generalization capacity against domain shift \cite{coupe_assemblynet_2020, kamraoui_deeplesionbrain_2022}. In this study, we propose a robust fusion strategy in the generation of the global deep grading map using validation accuracy. Our experiments show an improvement of model performance using this strategy. Moreover, this could also help to emphasize the brain locations related to AD, making the global deep grading map more reliable.

Finally, we propose to use graph-based modeling to better capture AD signature. Concretely, we propose to use graph convolutional network (GCN) model for AD classification problems. As a result, this shows state-of-the-art in performance for both AD diagnosis and prognosis.

This paper is an extension of the conference paper \cite{nguyen_deep_2021}, with several application-based contributions: (i) a study of graph design (\ie edge connectivity) and the choice of GCN as a classifier to boost the framework performance, (ii) an analysis of grading map interpretability with respect to the subject's age and (iii) a study of the consistency of our grading-based method to domain shift.

\section{Materials}

\subsection{Datasets}
The data used in this study, consisting of 2106 subjects, were obtained from multiple cohorts: the Alzheimer's Disease Neuroimaging Initiative (ADNI) \cite{jack_alzheimers_2008}, the Open Access Series of Imaging Studies (OASIS) \cite{lamontagne_oasis-3_2019}, the Australian Imaging, Biomarkers and Lifestyle (AIBL) \cite{ellis_australian_2009} and the Minimal Interval Resonance Imaging in Alzheimer's Disease (MIRIAD) \cite{malone_miriadpublic_2013}. We used the baseline T1-weighted MRI available in each of these studies. Each dataset contains AD patients and CN subjects. ADNI1 and AIBL datasets also include pMCI and sMCI patients. As in \cite{wen_convolutional_2020}, patients were considered as pMCI if they were diagnosed as MCI at the baseline and progressed to AD within 36 months. By contrast, patients were considered as sMCI if they were diagnosed as MCI at the baseline and all of sessions in the following 36 months. The group lists were obtained using ClinicaDL~\footnote{\url{https://github.com/aramis-lab/clinicadl}}
\cite{wen_convolutional_2020} and thus the selection criteria is similar. Table~\ref{tbl1} summarizes the number of participants and their age distribution for each dataset used in this study. T-tests showed no statistical differences in terms of age between two groups of the same dataset. During our experiments, AD and CN subjects from ADNI1 were used for training and all the other subjects as testing set. To minimize possible bias learned through training, we selected the same number of AD/CN subjects from ADNI1 for training without significant differences between the two age distributions ($p_{value}=0.27$). The evaluation consisted of two different tasks: Diagnosis (main task) and Prognosis (unknown task).

\begin{table*}[t]
\caption{Summary of participants used in our study. Data used for training are in bold}\label{tbl1}
\begin{tabular*}{0.75\textwidth}{@{\extracolsep{\fill}}lccccc}
\toprule
\textbf{Dataset} &  & \textbf{CN} & \textbf{AD} & \textbf{sMCI} & \textbf{pMCI} \\

\midrule
\multirow{3}{*}{ADNI1} & No. subjects & \textbf{170} & \textbf{170} & 129 & 171\\
& Age (Mean ± Std) & \textbf{75.9 ± 5.2} & \textbf{75.1 ± 7.2} & 74.6 ± 7.5 & 74.5 ± 7.0\\
& $p_{value}$ of t-test & \multicolumn{2}{c}{\textbf{0.27}} & \multicolumn{2}{c}{0.91}\\

\midrule
\multirow{3}{*}{ADNI2} & No. subjects & 149 & 181 &  & \\
& Age (Mean ± Std) & 74.1 ± 6.6 & 74.0 ± 7.2 &  & \\
& $p_{value}$ of t-test & \multicolumn{2}{c}{0.09} & \\

\midrule
\multirow{3}{*}{AIBL} & No. subjects & 232 & 47 & 12 & 30\\
& Age (Mean ± Std) & 72.3 ± 6.7 & 72.7 ± 8.6 & 72.5 ± 6.2 & 73.9 ± 8.0 \\
& $p_{value}$ of t-test & \multicolumn{2}{c}{0.72} & \multicolumn{2}{c}{0.67}\\

\midrule
\multirow{3}{*}{OASIS} & No. subjects & 658 & 98 &  & \\
& Age (Mean ± Std) & 68.6 ± 8.9 & 76.8 ± 8.4 &  & \\
& $p_{value}$ of t-test & \multicolumn{2}{c}{1.4e-16} & \\

\midrule
\multirow{3}{*}{MIRIAD} & No. subjects & 23 & 46 &  & \\
& Age (Mean ± Std) & 69.6 ± 7.0 & 69.3 ± 7.0 &  & \\
& $p_{value}$ of t-test & \multicolumn{2}{c}{0.86} & \\

\bottomrule
\end{tabular*}
\end{table*}
\subsection{Preprocessing}
All the T1w MRI were preprocessed using the following steps: (1) denoising \cite{manjon_adaptive_2010}, (2) inhomogeneity correction \cite{tustison_n4itk_2010}, (3) affine registration into MNI space ($181\times217\times181$ voxels at $1mm\times1mm\times1mm$) \cite{avants_reproducible_2011}, (4) intensity standardization \cite{manjon_robust_2008} and (5) intracranial cavity (ICC) extraction \cite{manjon_nonlocal_2014}. After preprocessing, we used AssemblyNet~\footnote{\url{https://github.com/volBrain/AssemblyNet}}  \cite{coupe_assemblynet_2020} to segment 133 brain structures (see Figure \ref{figure:pipeline}). In this study, brain structure segmentation is used to determine the structure volumes (\ie normalized volume in \% of ICC) and aggregate information to build the structured-based grading map (see Section \ref{section:deep_grading} and Figure \ref{figure:pipeline}).
\section{Method}

\subsection{Method overview}
\begin{figure*}[ht]
	\centering
		\includegraphics[width=0.8\textwidth]{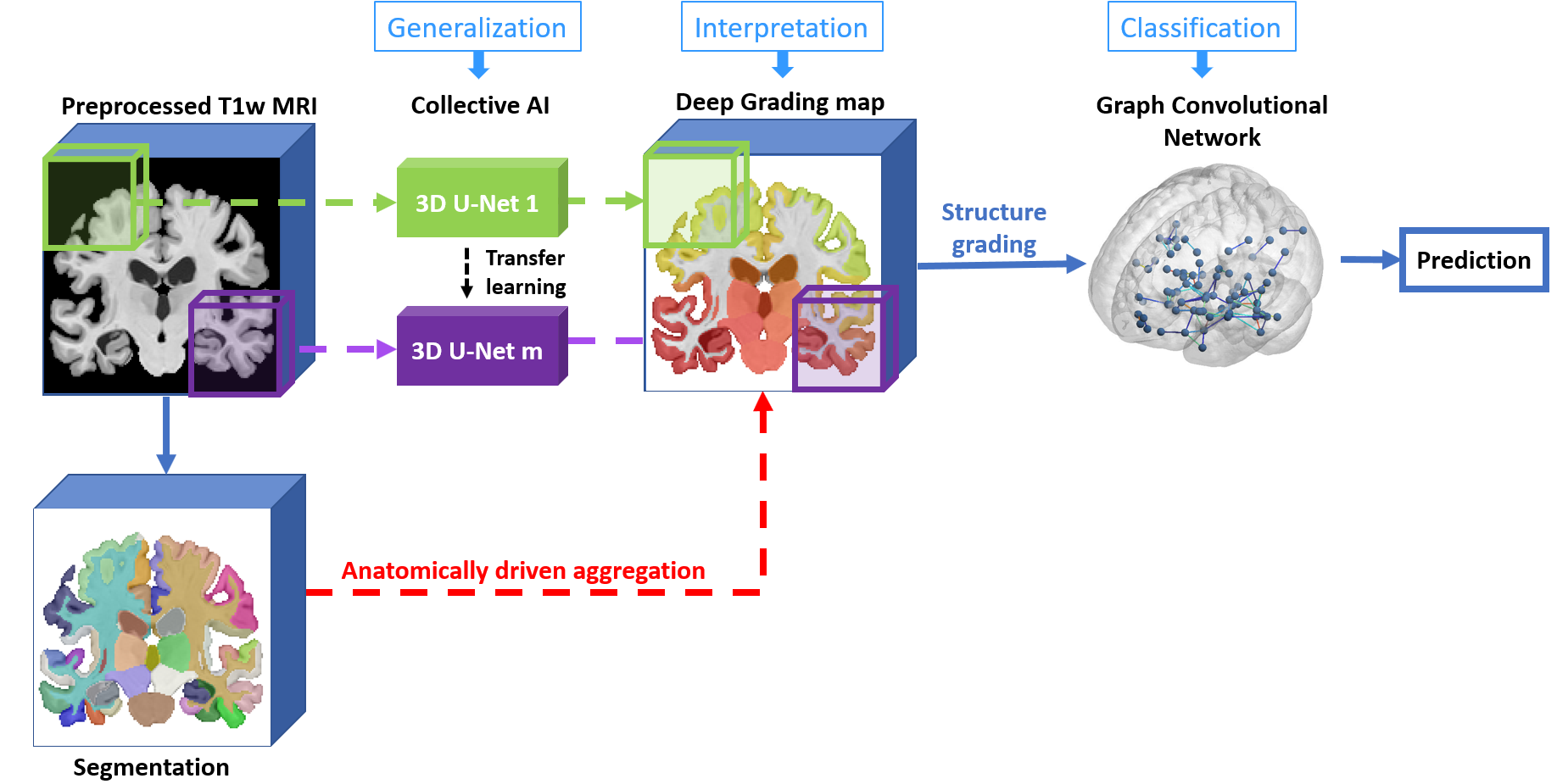}
	\caption{Overview of our processing pipeline. The MRI image, its segmentation and the deep grading map illustrated here are from an AD subject.}
	\label{figure:pipeline}
\end{figure*}

An overview of our proposed pipeline is shown in Figure \ref{figure:pipeline}. Our pipeline is designed based on different blocks, each of which serves a distinct purpose. First, the role of the collective AI block is to simulate a big model that cannot fit into a GPU by a large ensemble of smaller models. This strategy may help to capture more disease-related patterns than a single model. Indeed, it shows an improvement in generalization (see Section \ref{subsection:feature_study}) compared to other techniques. Second, the deep grading map provides a quantitative and interpretable assessment of the progression of AD. This 3D map can show AD-related regions, providing insight into the model prediction and helping clinicians in making reliable decisions. We use a segmentation here for a better visualization of the grading map and to reduce the data dimensionality in a meaningful way for experts. Finally, we use GCN to capture the relationship between brain structures. We demonstrate that GCN is well-adapted with grading features for AD detection (see Section \ref{subsection:classifier_comparison}).

Concretely, a preprocessed T1-weighted MRI with the size of $181 \times 217 \times 181$ voxels was downsampled to $91 \times 109 \times 91$ voxels to reduce the computational cost. The downsampled image was divided into $k \times k \times k$ (\ie $k=5$) overlapping patches of the same size (\ie $32 \times 48 \times 32$ voxels). We used $m=k\times k\times k$ (\ie $m=125$) 3D U-Nets to grade these patches. The 125 grading patches were fused to reconstruct a global grading map of $91 \times 109 \times 91$ voxels. This map was upscaled using interpolation to have the same size as the original input. After that, the segmentation of the original input (obtained with AssemblyNet \cite{coupe_assemblynet_2020}) was used to compute the average grading score for each structure. In this way, we obtained a vector of s elements where s is the number of segmented structures (\ie $s=133$). Finally, we created a fully connected graph with s nodes presenting the characteristic of s structures (\eg structure grading, structure volume, subject’s age) and used a graph convolutional neural network for the classification.
\subsection{Deep grading}
\label{section:deep_grading}
In AD diagnosis and prognosis, most of deep learning models only use CNN as a binary classification tool. In this study, we use CNN to produce 3D interpretable maps indicating the structural alterations caused by AD.

To capture these anatomical alterations, we extend the idea of several patch-based grading frameworks \cite{coupe_simultaneous_2012, tong_novel_2017, coupe_scoring_2012, hett_graph_2018}. The main objective is to provide a 3D grading map with a score between -1 and 1 at each voxel reflecting the disease severity. In \cite{coupe_simultaneous_2012}, the authors proposed to grade the hippocampus. For each voxel of this structure, they defined a surrounding patch and used a locally adaptive search algorithm to find the corresponding patch in all of training images. Similarity scores were then computed between the testing patch and training patches. These scores were used to estimate the grade (\ie degree of similarity to one group or another) for the considered voxel. Then, an average grading value was computed for the structure. The subject was classified as AD or CN depending on the sign of the grading value. They found that grading feature is more powerful than the measure of structure volume in distinguishing AD and CN subjects. Tong \ea used a sparse coding process to select a small number of discriminative voxels over the whole brain \cite{tong_novel_2017}. They showed that grading feature was efficient for AD prognosis even when training with AD/CN subjects. Contrary to these previous methods based on handcrafted feature extraction, here we propose a novel deep grading framework based on a large ensemble of 3D U-Nets (\ie 125 U-Nets).

Concretely, each of our 125 U-Nets (with the architecture similar to \cite{coupe_assemblynet_2020}) takes a 3D sMRI patch (\eg $32 \times 48 \times 32$ voxels in the MNI space) and outputs a grading map with value in range $[-1,1]$ for each voxel. Voxels with a higher value are considered closer to AD, while voxels with a lower value are considered closer to CN. For the ground-truth used during training, we assign the value 1 (resp. $-1$) to all voxels inside a patch extracted from an AD patient (resp. CN subject). All voxels outside of ICC are set to 0.

Once trained, the deep models are used to grade patches. These local outputs are gathered to reconstruct the final grading map (see Section \ref{section:collective_AI}). Using the structure segmentation, we represent each brain structure grading by its average grading score (see Figure \ref{figure:pipeline}). This anatomically driven aggregation allows better and meaningful visualization of the disease progression. In this way, during the classification step (see Section \ref{section:feature_classification}), each subject is encoded by an s-dimensional vector where s is the number of brain structures (\ie $s=133$).

\subsection{Collective AI}
\label{section:collective_AI}
In medical analysis, high generalization capacity across domains and unknown tasks presents potential clinical value as real data is diverse and may come from any source. As recently shown in \cite{bron_cross-cohort_2021, wen_convolutional_2020}, current deep learning methods for AD classification can well generalize to similar datasets but poorly perform on datasets having differences such as MRI protocols, age ranges, country of origin or inclusion criteria. In our testing datasets, different age range and MRI protocol were present in OASIS, different country of origin was present in AIBL and different inclusion criteria was present in MIRIAD. It should be noted that for OASIS, MCI and AD patients are mixed, so we used the ADNI inclusion criteria to separate AD patients and be able to assess the diagnosis of AD.

In this work, we propose to use an innovative collective artificial intelligence strategy to improve the generalization across domains and to unseen tasks. As recently shown for segmentation problems  \cite{coupe_assemblynet_2020, kamraoui_deeplesionbrain_2022}, the use of a large number of compact networks capable of communicating offers a better capacity for generalization. For brain segmentation, this strategy showed strong generalization to previously unexplored domains \cite{coupe_assemblynet_2020} (\ie trained on healthy adults and tested on children and AD patients). For the problem of multiple sclerosis lesion segmentation, this strategy also demonstrated the consistency across different natures of training domains \cite{kamraoui_deeplesionbrain_2022}. There are many other advantages of using the collective AI strategy. First, the use of a large number of compact networks is equivalent to a big neural network with more filters but the computation capacity required remains affordable. It should be noted that the same model taking the whole image at full resolution cannot be trained due to the limited memory of current GPUs. Second, the voting system based on a large number of specialized and diversified models helps the final grading decision to be more robust against domain shift and different tasks.

Concretely, after preprocessing and downsampling steps, we obtain $m=k \times k \times k$ patches $P_1,…,P_m$ (\ie $m=125$) with about 50\% overlapping volume. During training, for each patch location, a specialized model is trained. Therefore, we train $m$ 3D U-Nets to cover the whole image (see Figure \ref{figure:pipeline}). Moreover, each U-Net is initialized using transfer learning from its nearest neighbor U-Net, except the first one trained from scratch as proposed in \cite{coupe_assemblynet_2020}. As adjacent patches share common patterns, this communication allows grading models to share useful knowledge between them.

To obtain the final grading map, we propose a robust fusion strategy based on an average between overlapping patches, weighted by the accuracy obtained on the validation set. This weighted average for grading score fusion is computed as follows:

\begin{equation*}
    \label{eq:weighted_patch}
    G_i = \frac{\sum_{x_i \in P_j}{\alpha_j} * g_{ij}}{\sum_{x_i \in P_j}{\alpha_j}}
\end{equation*}

\noindent
where $G_i$ is the grading score of the voxel $x_i$ in the final grading map, $g_{ij}$ is the grading score of the voxel $x_i$ in the local grading patch $P_j$, and $\alpha_j$ is the balanced accuracy on validation of the patch $j$. This weighted vote enables to give more weight to the decision of accurate models during the reconstruction.
\subsection{Feature classification}
\label{section:feature_classification}
Most of current methods globally compared classes (\eg AD vs. CN) to perform classification. This kind of approach finds useful information from inter-subject similarities. For Alzheimer’s disease, the anatomical changes may occur in different brain areas and are different between subjects. These intra-subject variabilities may provide useful information for accurate AD detection. Consequently, it should be beneficial to combine these two characteristics for efficient classification. This can be done with the help of graph-based modeling. Indeed, following the idea of \cite{hett_graph_2018}, we modeled the intra-subject variabilities using a graph representation to capture the relationships between brain regions. We defined an undirected graph $\mathcal{G}=(\mathbb{N},\mathbb{E})$, where $\mathbb{N}={n_1,…,n_s}$ is the set of nodes for the $s$ brain structures and $\mathbb{E}=s \times s$ is the matrix of edge connections. In our approach, all nodes were connected with each other in a complete graph, where nodes embed brain features (\eg our proposed DG feature) and potentially other types of external features.

Indeed, besides the grading map, the volume of structures obtained from the segmentation could be helpful to distinguish AD patients from CN since AD yields to structure atrophy \cite{tong_novel_2017, hett_graph_2018}. In addition, the subject’s age is also an important factor since anatomical patterns in the brain of young AD patients could be similar to elder CN. Therefore, the combination of those features is expected to improve our classification performance. In our graph, each node could embed the structure grading score DG, structure volume V, and subject’s age A. All possible combinations are studied in Section \ref{subsection:feature_study}. Different types of graph edges are compared in Section \ref{subsection:edge_comparison}. Finally, we used a graph convolutional neural network (GCN) \cite{kipf_semi-supervised_2017} as the way to pass messages between nodes and perform the final decision. A comparison between different classifiers is provided in Section \ref{subsection:classifier_comparison} to explain our choice of GCN.
\subsection{Implementation details}
\label{section:implementation_details}

For each of the 125 patch locations, 80\% of the training dataset (\ie ADNI1) was used for training a 3D U-Net and the remaining 20\% for validation. To avoid bias resulting from dataset imbalance, the training/validation sets employed the same number of AD and CN. As the number of images in ADNI1 dataset was small, the training/validation data was re-split for each patch location to exploit the maximum information possible. The model was trained with voxel-wise mean absolute error (MAE) loss and Adam optimizer with a learning rate of 0.001. All voxels equally contribute to the loss function during training. The training process is stopped after 20 epochs without improvement in validation loss. We employed several data augmentation and sampling strategies to alleviate the overfitting issue during training. To train each U-Net, first, the corresponding cropping position of sub-volume was randomly translated by $t \in \{-1, 0, 1\}$ voxel in 3 dimensions of the image. Second, we sampled a sub-volume $X_1$ (with the label $Y_1$) from AD population, another sub-volume $X_2$ (with the label $Y_2$) from CN population and applied Mixup technique \cite{zhang_mixup_2018} to create a new sample: $X_{new} = \alpha X_1 + (1-\alpha) X_2$, $Y_{new} = \alpha Y_1 + (1-\alpha) Y_2$ where $\alpha \sim Beta(0.3, 0.3)$. This sample was used as the only input during the training.

Once the DG feature was obtained, we represented each subject by a graph of 133 nodes. Each node represented a brain structure and embeds its characteristic (\eg DG, V, A). Our classifier was composed of three layers of GCN \cite{kipf_semi-supervised_2017} with 32 channels, followed by a global mean average pooling layer and a fully connected layer with an output size of 1. The model was trained using the binary cross-entropy loss, Adam optimizer with a learning rate of 0.0003. No data augmentation was applied during training. The training process was stopped after 20 epochs without improvement in validation loss. During testing, we randomly added noise $X \sim \mathcal{N}(0,0.01)$ to the node features and computed the average of 3 predictions to get the global decision \cite{wang_test-time_2018}. Experiments showed that it helps our GCN to be more stable. For training and evaluating steps, we used a NVIDIA TITAN X with 12GB of memory. The total training time for $m=125$ U-Nets and the GCN model is about 23 hours. The total inference time of our method is about 1.63 seconds per preprocessed image.
\section{Experimental results}

\subsection{Performance study}
In this section, the 125 CNN grading models and the classifier were trained using AD and CN subjects of the ADNI1 dataset. Then, we assessed their generalization capacity to domain shift using AD and CN subjects from ADNI2, AIBL, OASIS and MIRIAD. The generalization capacity for unseen tasks was studied using pMCI, sMCI subjects (AD prognosis) from ADNI1 (same domain) and AIBL (out of domain). Due to the imbalanced nature of testing datasets, we used the balanced accuracy (BACC) and area under receiver operating characteristic curve (AUC) to measure the performance of different classifiers. The global BACC/AUC for diagnosis and prognosis was measured with all available testing images for each task. Each experiment was repeated ten times (to reduce bias related to random nature of DL training) and the average results was provided as final results. All comparisons were made using the Wilcoxon test by comparing the ten BACC/AUC values obtained over the 10 repetitions as recommended in \cite{demsar_statistical_2006}. The one-sided test was applied to confirm a superior performance. A confidence level of $5\%$ is used so that $p_{value}<0.05$ means the considered result is significantly better than a chosen baseline.
\subsubsection{Features for classification}
\label{subsection:feature_study}

In this part, we study the different feature types used as input of the final classifier. The edges connecting the graph nodes are set to 1 in this comparison. The DG feature is denoted as $DG_C$ (resp. $DG_I$) when obtained with the collective (resp. individual) AI strategy. The individual AI strategy refers to the use of a single U-Net to learn patterns from all patches of the input image. We also denote $DG_{Cnw}$ for the no-weighted version of $DG_C$. The results of BACC performance are presented in Table \ref{table2}. The result of AUC are in annexes (Table \ref{table5_auc}).

\begin{table*}[t]
\caption{Comparison of different types of features for classification. All the edges are set to 1, the classifier used is GCN. {\color{red} Red}: best result, {\color{blue} Blue}: second best result. The balanced accuracy (BACC) is used to assess the model performance. The results are the average accuracy of 10 repetitions and presented in percentage. All the methods were trained on the AD/CN subjects of the ADNI1 dataset. Value in bold: $p$ of one-sided Wilcoxon test comparing with our baseline (in gray) is lower than 0.05, meaning a significantly superior performance is found compared to the baseline. A comparison using area under curve (AUC) is provided in annexes.}\label{table2}
\begin{tabular*}{0.93\textwidth}{@{\extracolsep{\fill}}cccccccccc}
\toprule
\multirow{3}{*}{No.} & \multirow{3}{*}{Features} & \multicolumn{4}{c}{\makecell{Diagnosis \\ (AD/CN)}} & \multicolumn{2}{c}{\makecell{Prognosis \\ (p/sMCI)}} & \makecell{Global \\ Diagnosis \\ (AD/CN)} & \makecell{Global \\ Prognosis \\ (p/sMCI)} \\
& & ADNI2 & AIBL & OASIS & MIRIAD & ADNI1 & AIBL & All & All\\
& & $N=330$ & $N=279$ & $N=756$ & $N=69$ & $N=300$ & $N=32$ & $N=1434$ & $N=332$\\

\midrule

1 &  $DG_I$ & {\color{red} \textbf{88.6}} & 82.3 & 88.0 & 96.2 & 68.2 & 71.4 & 88.4 & 68.2\\
2 &  $DG_{Cnw}$ & 86.4 & 88.0 & {\color{red} 89.1} & {\color{blue} 99.3} & 70.3 & 73.0 & 88.5 & 70.4\\
3 &  $DG_C$ & 87.2 & 88.5 & {\color{blue} 88.9} & {\color{red} 99.8} & 70.6 & {\color{red} 75.4} & {\color{blue} 89.0} & 71.0\\
4 &  $V$    & 67.4 & 64.0 & 72.8 & 70.6 & 56.1 & 61.2 & 69.8 & 56.5\\
5 &  $A$    & 50.5 & 52.7 & 46.1 & 42.2 & 49.8 & 50.3 & 46.5 & 50.0\\
6 &  $V, A$ & 63.2 & 59.8 & 58.5 & 54.5 & 52.9 & 55.7 & 57.6 & 53.0\\
7 &  $DG_C, V$ & 86.3 & 88.4 & 88.4 & 98.7 & 70.8 & {\color{blue} 75.2} & 88.3& {\color{blue} 71.0}\\
\rowcolor{Gray}
8 &  $DG_C, A$ & {\color{blue} 87.5} & {\color{red} 92.1} & 88.8 & 99.0 & {\color{blue} 73.8} & 74.5 & {\color{red} 89.5} & {\color{red} 73.7}\\
9 &  $DG_C, V, A$ & 87.3 & {\color{blue} 91.8} & 88.2 & 98.7 & {\color{red} 73.9} & 72.7 & 88.9 & {\color{blue} 73.6}\\

\bottomrule
\end{tabular*}
\end{table*}

\subsubsection*{Comparison of Grading vs. Volume}

As discussed previously, brain atrophy is an important biomarker of Alzheimer’s disease. Many studies used structure volume for AD classification and achieved encouraging results \cite{guo_grey-matter_2014, ledig_structural_2018, schmitter_evaluation_2015}. So, we compare the proposed biomarker (grading, exp. 3) and the classical one (volume, exp. 4) to assess the efficiency of our new biomarker. The additional evaluation using the age feature (exp. 5) was performed to confirm that no age bias was present in the training/testing partitions.

The efficiency of $DG_C$ (exp. 3) was clearly better than V (exp. 4).  $DG_C$ outperformed V in global diagnosis, global prognosis and all of the tests on an individual dataset (all $p_{value}<0.05$). Thus, the proposed biomarker $DG_C$ presents an important interest for AD classification.

\begin{figure*}[ht]
	\centering
		\includegraphics[width=0.9\textwidth]{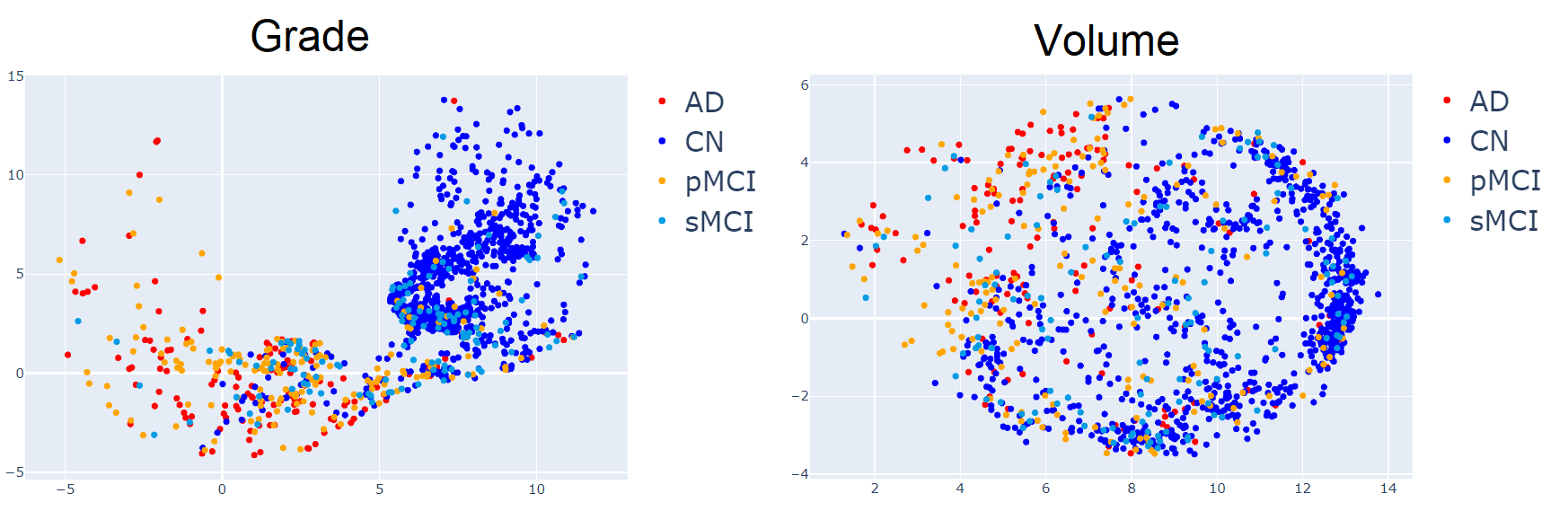}
	\caption{UMAP visualization of test set.}
	\label{figure:umap}
\end{figure*}

Moreover, we trained a UMAP \cite{mcinnes_umap_2020} with AD/CN subjects from ADNI1 and visualized the transformed test set in 2D space (see Figure \ref{figure:umap}). The transformed data was colored with respect to the diagnosis class. Two types of input were considered: grade ($DG_C$ ) and volume (V). The grading feature was visually better to separate AD and CN subjects than the volume feature. To confirm this assessment, we applied K-means with 2 clusters (we considered 1 cluster for CN/sMCI and 1 cluster for AD/pMCI) to this 2D data to assess the separability of the two clusters. The silhouette score \cite{rousseeuw_silhouettes_1987} was used to measure this separability. This score ranges from $-1$ to 1. A higher value means clusters are more distinguishable. As a result, the silhouette score obtained with $DG_C$ was 0.55, better than 0.41 obtained with V.

\subsubsection*{Comparison of Collective AI features vs. Individual AI features}

We aimed at assessing the efficiency of the collective AI strategy. To do this, we compared the efficiency of $DG_C$ and $DG_I$ features (exp. 1, 3) (see Table \ref{table2}). Experimental results showed that $DG_C$ (exp. 3) is significantly better than $DG_I$ (exp. 1) for both global diagnosis  ($p_{value}=0.007$) and global prognosis ($p_{value}=0.001$). Consequently, collective AI strategy offered a significant improvement for unseen domain (AD diagnosis) and unseen task (AD prognosis). In terms of generalization, $DG_C$ alleviated the drop in performance in AIBL dataset for AD diagnosis.

\subsubsection*{Efficiency of the weighted fusion strategy in Collective AI features}
We validate the efficiency of the weighted fusion strategy in Collective AI by comparing this strategy with its no-weighted version (exp. 2, 3) (see Table \ref{table2}). Experimental results showed that $DG_C$ (exp. 3) is significantly better than $DG_{Cnw}$ (exp. 2) for global diagnosis  ($p_{value}=0.019$) and similar for global prognosis ($p_{value}=0.188$). Consequently, the weighted fusion strategy can improve the model performance in AD diagnosis while keeping a good performance in AD prognosis.

\subsubsection*{Combination of grading and additional features}

Several works showed that complementary information about the subject could help to improve the performance of their classifier \cite{tong_novel_2017, hett_multi-scale_2021}. In these studies, different cognitive scores were used such as MMSE, CDR-SB, RAVLT, FAQ, ADAS11, and ADAS13 cognitive tests. However, this information is not always available. Instead, we employed brain structure volumes and the subject’s age as additional features here.

Four experiments were made using multiple types of features in graph nodes (exp. 6, 7, 8, 9). The best performance of diagnosis and prognosis was obtained using $DG_C,A$ (exp. 8) and it was significantly better than using only $DG_C$ (exp. 3) for both global scores (all $p_{value}<0.05$). Overall, using subject’s age in addition to $DG_C$ produced the best results. Consequently, in the rest of the paper, we use $DG_C$ and the age feature as input for further analysis.

\subsubsection{Comparison of different types of graph edges}
\label{subsection:edge_comparison}
In this part, we compare different types of graph edges. In general, when constructing a graph from neuroimage data, there are different ways to define the connection between nodes \cite{bessadok_graph_2021}. Huang et al. defined it by counting fiber tracts in Diffusion Tensor Imaging \cite{huang_attention-diffusion-bilinear_2020}. Li et al. computed this as the pairwise correlations of functional magnetic resonance imaging time series \cite{li_braingnn_2020}. With sMRI data, Mahjoub et al. defined the connection between two ROI as the absolute difference between their averaged cortical attributes \cite{mahjoub_brain_2018}. In this study, we propose different edge types as follows: Fully-one edge (all edges are set to 1), correlation-based edge (the edge connecting each pair of brain structures is defined as the Pearson's correlation based on their grading scores), volume difference-based edge (the edge connecting each pair of brain structures is the absolute difference of their volumes). The results of the comparison are presented in Table \ref{table:edges_comparison}. We observe that the edge based on structure volume difference leads to a better classification performance than other tested types of edge in all datasets and all tasks. Thus, we use the edge based on structure volume difference in the rest of the paper.

\begin{table*}[t]
\caption{Comparison of different graph edge types. The classifier used is GCN and the input features is $DG_C$ and A. {\color{red} Red}: best result, {\color{blue} Blue}: second best result. The balanced accuracy (BACC) is used to assess the model performance. The results are the average accuracy of 10 repetitions and presented in percentage. All the methods were trained on the AD/CN subjects of the ADNI1 dataset. A comparison using area under curve (AUC) is provided in annexes.}\label{table:edges_comparison} 
\begin{tabular*}{0.97\textwidth}{@{\extracolsep{\fill}}ccccccccc}
\toprule
\multirow{3}{*}{Edge} & \multicolumn{4}{c}{\makecell{Diagnosis \\ (AD/CN)}} & \multicolumn{2}{c}{\makecell{Prognosis \\ (p/sMCI)}} & \makecell{Global \\ Diagnosis \\ (AD/CN)} & \makecell{Global \\ Prognosis \\ (p/sMCI)} \\
& ADNI2 & AIBL & OASIS & MIRIAD & ADNI1 & AIBL & All & All\\
& $N=330$ & $N=279$ & $N=756$ & $N=69$ & $N=300$ & $N=32$ & $N=1434$ & $N=332$\\

\midrule
Fully-one & {\color{blue} 87.5} & {\color{blue} 92.1} & {\color{blue} 88.8} & {\color{blue} 99.0} & {\color{blue} 73.8} & {\color{blue} 74.5} & {\color{blue} 89.5} & {\color{blue} 73.7}\\
Correlation & {\color{blue} 87.5}	& 91.8	& 88.4	& 98.6	& 73.4	& 74.1	& 89.2	& 73.3\\
Volume difference & {\color{red} 87.6} & {\color{red} 92.4} & {\color{red} 89.1} & {\color{red} 99.6} & {\color{red} 73.9} & {\color{red} 75.6} & {\color{red} 89.6} & {\color{red} 73.9}\\

\bottomrule
\end{tabular*}
\end{table*}

\subsubsection{Comparison of different classifiers}
\label{subsection:classifier_comparison}
In this section, we study different solutions for the graph classification. We compare the use of GCN with other classifiers such as SVM, multi-layer perceptron, Transformer Graph \cite{shi_masked_2020}, sample and aggregate graph (SAGE) \cite{hamilton_inductive_2017}, residual gated graph (ResGatedGraph) \cite{bresson_residual_2017},  graph attention network (GAT) \cite{velickovic_graph_2017} and topology adaptive graph (TAG)~\cite{du_topology_2017}. Table \ref{table:architecture_comparison} shows the results of this comparison. We can observe that GCN achieves the best performance most of the time. Consequently, we chose GCN as a classifier in our framework.

\begin{table*}[t]
\caption{Comparison of different classifiers. For graph-based approaches (\ie all the approaches except SVM and multi-layer perceptron), the edge based on structure volume difference is used and the input features is $DG_C$ and A. {\color{red} Red}: best result, {\color{blue} Blue}: second best result. The balanced accuracy (BACC) is used to assess the model performance. The results are the average accuracy of 10 repetitions and presented in percentage. All the methods were trained on the AD/CN subjects of the ADNI1 dataset. A comparison using area under curve (AUC) is provided in annexes.}\label{table:architecture_comparison}
\begin{tabular*}{0.97\textwidth}{@{\extracolsep{\fill}}ccccccccc}
\toprule
\multirow{3}{*}{Classifier} & \multicolumn{4}{c}{\makecell{Diagnosis \\ (AD/CN)}} & \multicolumn{2}{c}{\makecell{Prognosis \\ (p/sMCI)}} & \makecell{Global \\ Diagnosis \\ (AD/CN)} & \makecell{Global \\ Prognosis \\ (p/sMCI)} \\
& ADNI2 & AIBL & OASIS & MIRIAD & ADNI1 & AIBL & All & All\\
& $N=330$ & $N=279$ & $N=756$ & $N=69$ & $N=300$ & $N=32$ & $N=1434$ & $N=332$\\

\midrule
SVM & 85.7 & 88.7 & 87.4 & 95.6 & 69.0 & 69.7 & 87.6 & 68.9\\
Multi-layer perceptron & 82.5 & 87.4 & 83.4 & 88.0 & 66.4 & 61.7 & 84.6 & 65.8\\
Transformer & 87.9 & 91.3 & 87.9 & {\color{blue} 98.5} & 72.8 & {\color{blue} 75.4} & 89.1 & 72.9\\
SAGE & 87.2 & {\color{blue} 91.8} & 88.1 & 98.3 & {\color{blue} 73.4} & 73.3 & 88.9 & {\color{blue} 73.2}\\
ResGatedGraph & 84.6 & 87.6 & 81.9 & 92.7 & 72.5 & 70.8 & 84.0 & 70.3\\
GAT & {\color{red} 87.7} & 91.6 & {\color{blue} 88.7} & 98.2 & {\color{blue} 73.4} & 72.5 & {\color{blue} 89.3} & 73.1\\
TAG & 87.4 & 91.3 & 87.8 & 97.7 & 73.3 & 74.2 & 88.8 & {\color{blue} 73.2}\\
GCN & {\color{blue} 87.6} & {\color{red} 92.4} & {\color{red} 89.1} & {\color{red} 99.6} & {\color{red} 73.9} & {\color{red} 75.6} & {\color{red} 89.6} & {\color{red} 73.9}\\

\bottomrule
\end{tabular*}
\end{table*}

\subsubsection{Comparison with state-of-the-art methods}
Tables \ref{table3} and \ref{table4} summarize the current performance in BACC of state-of-the-art methods proposed for AD diagnosis and prognosis classification that have been validated on external datasets. A comparison of performance in AUC is provided in annexes (Tables \ref{table6_auc} and \ref{table7_auc}). In this comparison we consider five categories of deep methods: patch-based strategy based on a single model (Patch-based CNN \cite{wen_convolutional_2020}), patch-based strategy based on multiple models (Landmark-based CNN \cite{liu_landmark-based_2018}, Hierarchical FCN \cite{lian_hierarchical_2020}), ROI-based strategy based on a single model focused on hippocampus (ROI-based CNN \cite{wen_convolutional_2020}), subject-based considering the whole image based on a single model (Subject-based CNN \cite{wen_convolutional_2020}, Efficient 3D \cite{yee_construction_2021} and  $AD^2A$ \cite{albarqouni_attention-guided_2020}) and a classical voxel-based model using a SVM (Voxel-based SVM \cite{wen_convolutional_2020}). Only methods evaluated across different datasets were selected here.

\subsubsection*{Comparison with methods under the same condition}
For a fair comparison, we retrained and evaluated four methods whose code is available: Patch-based CNN, ROI-based CNN, Subject-based CNN and Voxel-based SVM \cite{wen_convolutional_2020} with our training/testing data. The results are reported in Table~\ref{table3}.

For AD diagnosis (\ie AD/CN), as ADNI2 and ADNI1 (training set) are very similar, we used the performance on ADNI2 as a reference to assess the capacity of generalization on other datasets (\ie AIBL, OASIS, MIRIAD). Based on that, we observed a major drop in performance in Patch-based CNN method for AIBL, OASIS and MIRIAD, ROI-based CNN method for AIBL (see Table \ref{table3}). For AD prognosis (\ie pMCI/sMCI), we also observed a drop in performance between AIBL and ADNI1 (training domain) in Patch-based CNN method, ROI-based CNN method and Subject-based CNN method. Overall, our method shows a good generalization capacity against domain shift and to unseen tasks compared to other methods. Moreover, our method always achieves the best result in terms of performance for all datasets/tasks and outperforms the traditional method (\ie Voxel-based SVM) by a large margin.

\begin{table*}[t]
\caption{Comparison of our method with state-of-the-art methods with available code that have been retrained on our training dataset and tested on our dataset. {\color{red} Red}: best result, {\color{blue} Blue}: second best result. The balanced accuracy (BACC) is used to assess the model performance. All the methods are trained on the AD/CN subject of the ADNI1 dataset, the same training/testing partition is used for evaluation. A comparison using area under curve (AUC) is provided in annexes.}\label{table3}
\begin{tabular*}{\textwidth}{@{\extracolsep{\fill}}lcccccc}
\toprule
\multirow{2}{*}{Method} & \multicolumn{4}{c}{\makecell{Diagnosis \\ (AD/CN)}} & \multicolumn{2}{c}{\makecell{Prognosis \\ (p/sMCI)}}\\
& \makecell{ADNI2 \\ $N=330$}& \makecell{AIBL \\ $N=279$}& \makecell{OASIS \\ $N=756$}& \makecell{MIRIAD \\ $N=69$}& \makecell{ADNI1 \\ $N=300$}& \makecell{AIBL \\ $N=32$}\\

\midrule

Patch-based CNN \cite{wen_convolutional_2020} & 72.4 & 63.4 & 67.5 & 63.0 & 62.5 & 47.5\\
ROI-based CNN \cite{wen_convolutional_2020} & 79.7 & 74.4 & 79.0 & 81.5 & 65.5 & 62.5\\
Subject-based CNN \cite{wen_convolutional_2020} & 76.1 & 81.5 & 86.0 & 89.1 & 64.8 & 55.8\\
Voxel-based SVM \cite{wen_convolutional_2020} & {\color{blue} 83.3} & {\color{blue} 88.2} & {\color{blue} 87.4} & {\color{blue} 93.5} & {\color{blue} 67.2} & {\color{blue} 70.0}\\

\midrule
Our method & {\color{red} 87.6} & {\color{red} 92.4} & {\color{red} 89.1} & {\color{red} 99.6} & {\color{red} 73.9} & {\color{red} 75.6}\\

\bottomrule
\end{tabular*}
\end{table*}

\subsubsection*{Literature comparison}
We also detail the results of four other methods without available implementation performing evaluation across different datasets. In this case, we present the results of the original papers in Table~\ref{table4}. Consequently, there are many different factors between methods: number of subjects in training/testing sets, selection criteria, etc. However, this could help to get an idea of the performance of current methods in the application of AD diagnosis/prognosis.

Overall, our method has most of the time the best or the second best result. Furthermore, it should be noted that our model is trained using only 340 images (from ADNI1) without any domain adaptation technique but outperforms Efficient3D (trained on 2843 images) and $AD^2A$ (with domain adaptation) in most of datasets/tasks.

\begin{table*}[t]
\caption{Comparison of our method with state-of-the-art methods using published results. {\color{red} Red}: best result, {\color{blue} Blue}: second best result. The balanced accuracy (BACC) is used to assess the model performance. All the methods are trained on the AD/CN subject of the ADNI1 dataset. However, there are many different factors: number of subjects in training/testing sets, selection criteria, etc. A comparison using area under curve (AUC) is provided in annexes.}\label{table4}
\begin{tabular*}{\textwidth}{@{\extracolsep{\fill}}lcccccc}
\toprule
\multirow{2}{*}{Method} & \multicolumn{4}{c}{\makecell{Diagnosis \\ (AD/CN)}} & \multicolumn{2}{c}{\makecell{Prognosis \\ (p/sMCI)}}\\
& ADNI2 & AIBL & OASIS & MIRIAD & ADNI1 & AIBL \\

\midrule

Landmark-based CNN \cite{liu_landmark-based_2018} & {\color{red} 90.8} & - & - & 92.4 & - & -\\
Hierarchical FCN \cite{lian_hierarchical_2020} & {\color{blue} 89.5} & - & - & - & 69.0 & -\\
$AD^2A$ \cite{albarqouni_attention-guided_2020} & 88.3 & 87.8 & - & - & - & -\\
Efficient3D \cite{yee_construction_2021} & - & {\color{blue} 90.7} & {\color{red} 91.9} & {\color{blue} 95.7} & {\color{blue} 70.1} & {\color{blue} 65.2}\\

\midrule
Our method & 87.6 & {\color{red} 92.4} & {\color{blue} 89.1} & {\color{red} 99.6} & {\color{red} 73.9} & {\color{red} 75.6}\\

\bottomrule
\end{tabular*}
\end{table*}

\subsection{Interpretation of deep grading maps}
\label{section:interpretation_dg_maps}
\begin{figure*}[ht]
	\centering
		\includegraphics[width=0.7\textwidth]{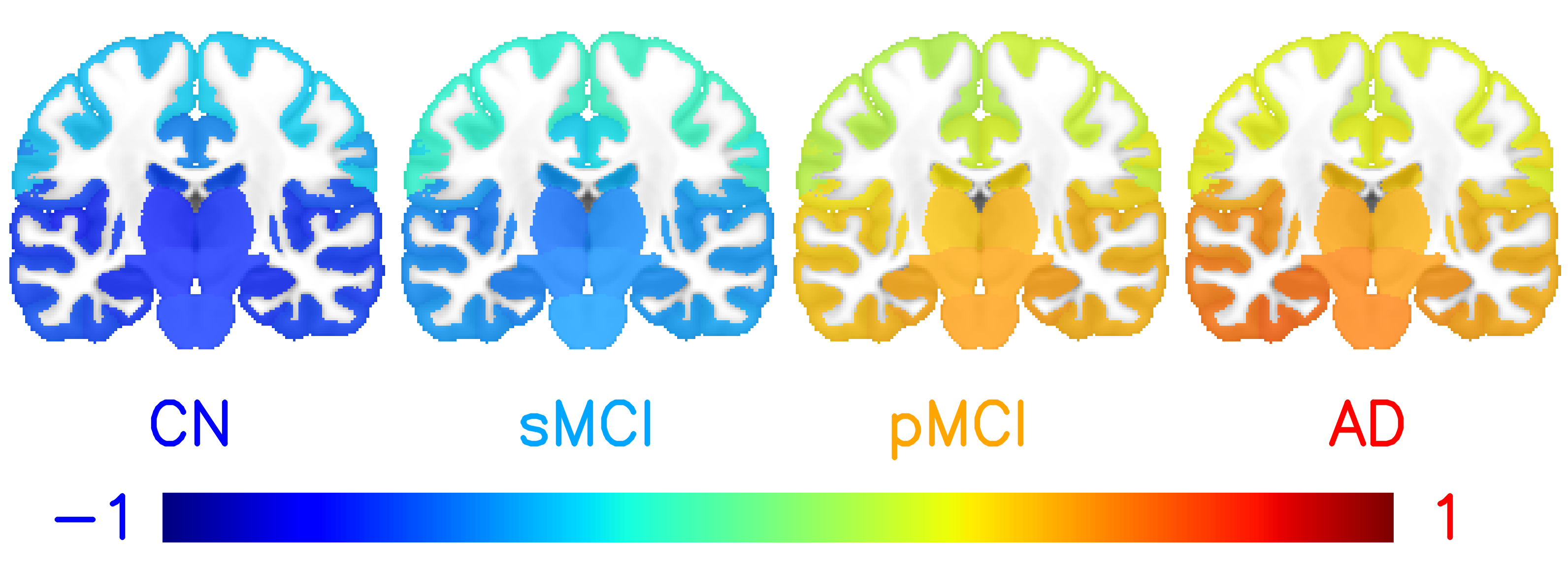}
	\caption{Average grading map per group of subjects.}
	\label{figure:avg_grading_map}
\end{figure*}

To highlight the interpretability capabilities offered by our DG feature, we computed the average DG map for each group: AD, pMCI, sMCI and CN (see Figure \ref{figure:avg_grading_map}). First, we could note that the average grading increased between each stage of the disease. Second, we estimated the top 10 structures with highest absolute value of grading score over all the testing subjects. Nine of these structures were known to be specifically and early impacted by AD. These structures were: bilateral hippocampus \cite{frisoni_clinical_2010}, left amygdala and left inferior lateral ventricle \cite{coupe_lifespan_2019}, left parahippocampal gyrus \cite{kesslak_quantification_1991}, left posterior insula \cite{foundas_atrophy_1997}, left thalamus \cite{de_jong_strongly_2008}, left transverse temporal gyrus \cite{addneuromed_consortium_education_2012}, left ventral diencephalon \cite{lebedeva_mri-based_2017}. These results showed a high correlation with current physiopathological knowledge on AD \cite{jack_hypothetical_2010}.

\begin{figure*}[ht]
	\centering
		\includegraphics[width=0.6\textwidth]{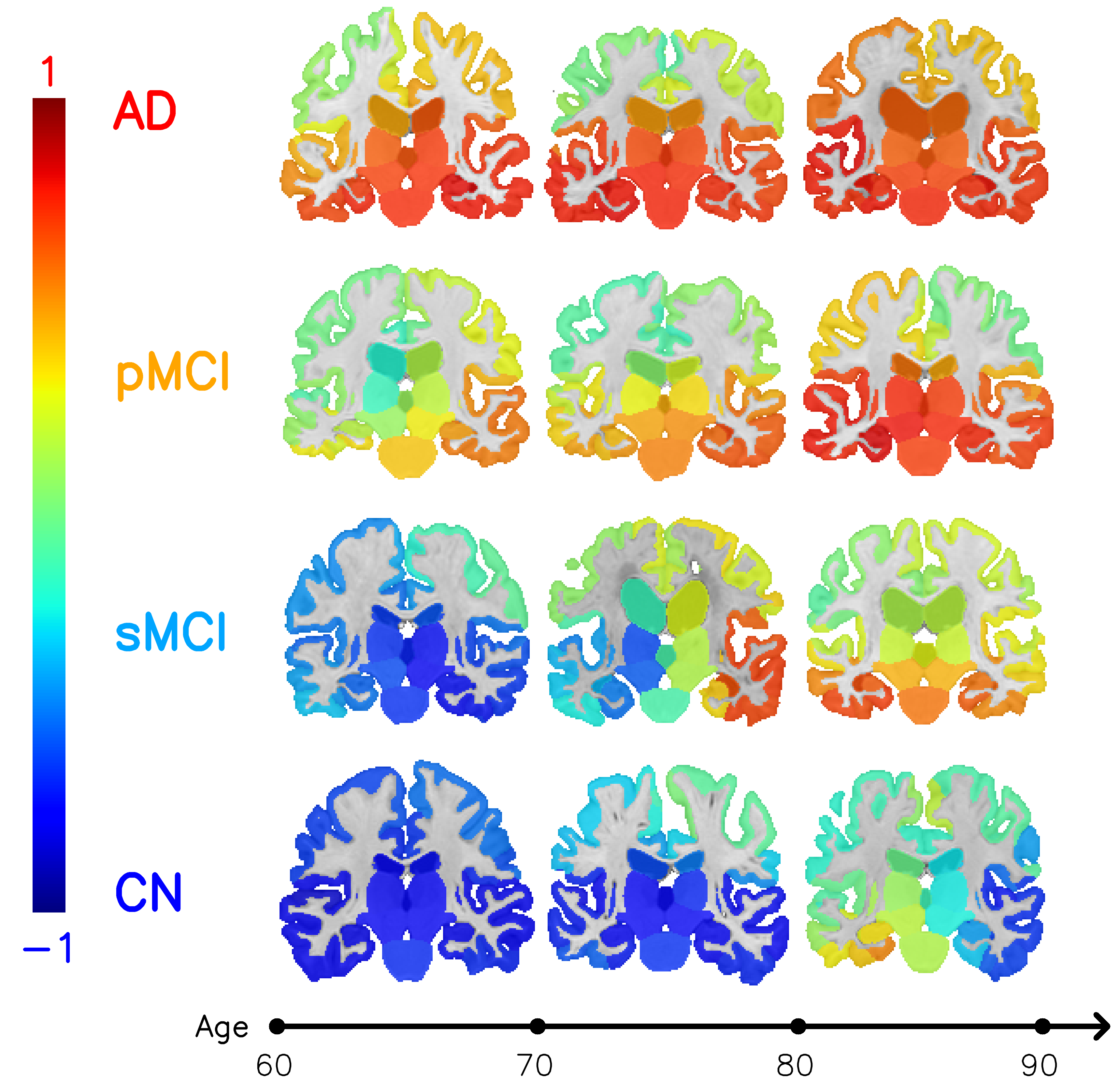}
	\caption{Typical grading maps (from individual subjects) for each state of disease with respect to age.}
	\label{figure:individual_grading_map}
\end{figure*}

Typical individual grading maps of each population (\ie CN, sMCI, pMCI, AD) were selected and are presented in Figure~\ref{figure:individual_grading_map}. First, we observed that older people had higher grade than younger people as expected. Second, for the same age range, the color of grading maps changed progressively depending to the disease severity. Third, CN/AD populations seemed to be more distinguishable from each other than sMCI/pMCI populations. We observed high similarity between older sMCI patients (80-90 years old) and younger pMCI patients (60-70 years old). This might be the reason why the performance of AD prognosis was lower than AD diagnosis and why the use of age improved the results of AD prognosis. Finally, we observed that the earliest brain alteration started from hippocampus and its surrounding regions (sMCI at 70-80 years old in Figure~\ref{figure:individual_grading_map}) and spanned over time to the whole brain (AD at 80-90 years-old in Figure~\ref{figure:individual_grading_map}). All of these findings demonstrated the potential capacity of deep grading maps to assist clinicians in practice.

\subsection{Consistency study}
Thibeau-sutre \ea have recently shown that for the same CNN architecture, different training data or even training runs can lead to different explanations \cite{thibeau-sutre_visualization_2020}.  They suggested that a good explanation method should not depend on training data or training initialization. In this study, we analyzed these two aspects for our grading maps (dependency to data training and model initialization) by performing two experiments. 

First, we trained two grading models (each one consisting of 125 U-Nets) on ADNI1 (model 1) and ADNI2 (model 2) datasets. For each model, we then calculated the $DG_C$ vector from the grading map for all images in testing set (excluding ADNI1 and ADNI2). Finally, we measured the cosine similarity of two $DG_C$ vectors obtained from each image. We obtained a median of 0.92 as similarity between two $DG_C$ vectors from two models training on different datasets that demonstrate the good robustness to domain shift of our method.

For the second one, we trained the grading model twice using only ADNI1 as training set (models 1 \& 3). Finally, we obtained a median of 0.95 as similarity between $DG_C$ vectors from two retrained models on the same dataset that demonstrate the good robustness to training initialization of our method.

\begin{figure*}[ht]
	\centering
		\includegraphics[width=0.8\textwidth]{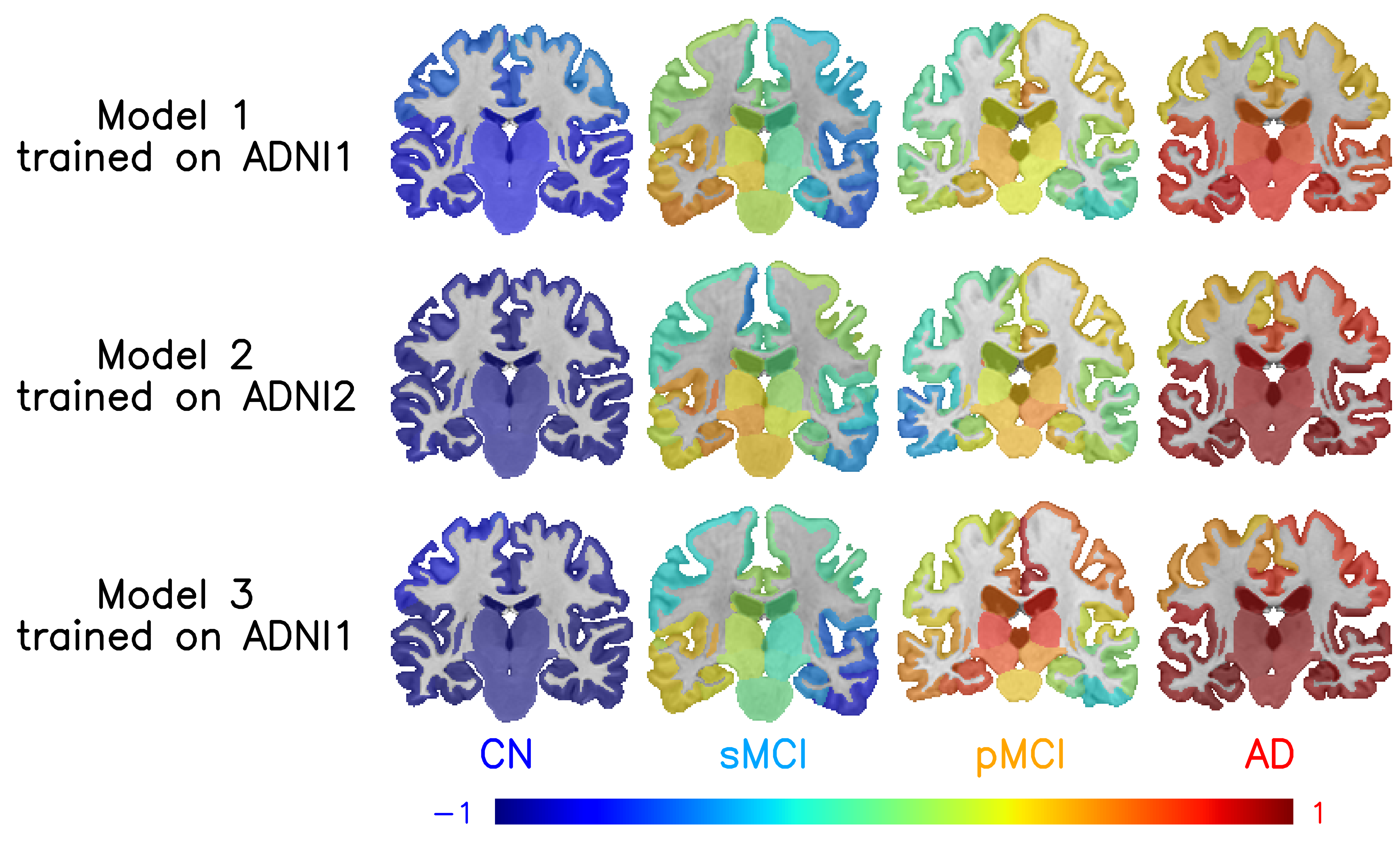}
	\caption{Consistency of grading maps between retrained grading models (models 1 \& 3), and between grading models trained on different datasets (models 1 \& 2).}
	\label{figure:consistancy}
\end{figure*}

Figure \ref{figure:consistancy} shows examples of individual grading maps for four considered populations (\ie CN, sMCI, pMCI, AD). We can visually see the similarities between grading models trained on different datasets (\ie models 1 \& 2) and between grading models trained several times on the same dataset (\ie models 1 \& 3). Overall, the three models identify AD-related areas in a similar way. These experiments show the consistency of Deep Grading maps across different training runs and different training sets.

\section{Discussion}
In this paper, we proposed a novel deep grading framework dedicated to Alzheimer's disease diagnosis and prognosis. Our framework was designed to overcome three main limitations of current deep learning methods for AD classification: performance compared to conventional machine learning method (\ie SVM), generalization to unseen datasets/tasks and interpretability.

While many studies found that deep learning and SVM methods had a similar performance for AD classification problem \cite{bron_cross-cohort_2021, wen_convolutional_2020}, the authors also suggested that better model design could improve the performance of DL methods. Indeed, Jo \ea indicate that hybrid methods using CNN features and a conventional classifier showed better accuracy than pure deep learning methods \cite{jo_deep_2019}. In this study, we combined CNN features with a GCN classifier. The use of a GCN also allows to combine additional demographic information to further improve the model performance. As a result, our model showed a better performance with a large margin compared to traditional methods (\eg SVM). Furthermore, a careful design of edge connectivity in the subject's graph may also boost the model performance. In this study, we propose to define the edge connection between two brain structures as the absolute difference of their volumes. This type of connectivity allows our model to make more accurate decisions. The analysis of this connectivity is provided in annexes.

This study is one of the few assessing deep model performance on multiple independent datasets (\ie ADNI2, AIBL, OASIS, MIRIAD) and unseen tasks (\ie AD prognosis) \cite{bron_cross-cohort_2021, wen_convolutional_2020, yee_construction_2021, liu_landmark-based_2018, lian_hierarchical_2020, albarqouni_attention-guided_2020}. For AD diagnosis, the result on ADNI2 dataset was 87.6\% in BACC which was competitive with the current performance reported in the literature. On OASIS, we achieved the second place with 89.1\% accuracy. On AIBL and MIRIAD, our model outperformed current state-of-the-art methods with respectively 92.4\% and 99.6\%. Besides several studies found a drop in performance when evaluating on independent datasets \cite{bron_cross-cohort_2021, wen_convolutional_2020}. This performance drop could come from differences in MRI protocols, age ranges, country of origin and inclusion criteria. Our results demonstrated the high generalization capacity of our method against datasets with such differences. Especially, the use of collective AI enabled a better generalization to unseen tasks (\ie AD prognosis) than other deep learning methods. Finally, the use of the weighted fusion strategy could improve even more the model performance (in AD diagnosis).

In terms of interpretability, our framework provides 3D grading maps capable of indicating regions impacted by AD. The most important structures highlighted by our grading map were correlated with knowledge about the disease in the literature. Furthermore, our experiments showed that grading features were more efficient than volume features for both AD diagnosis and prognosis which confirmed the finding of \cite{coupe_simultaneous_2012, coupe_detection_2015}. When coupling the grading map with a GCN classifier, it yielded high performance across datasets. Hence, grading maps are not only an interpretable visualization but provide also discriminative features for AD classification. However, the use of GCN made our framework become not fully interpretable. The fully-interpretable framework can be done by replacing the final GCN by an SVM classifier or a simple threshold. Although, this implies a trade-off between interpretability and performance.

Compared to the explanation maps of explainable methods such as Class Activation Mapping \cite{zhou_learning_2015} and Layer-wise relevance propagation \cite{binder_layer-wise_2016}, our grading map exhibits interesting properties. Indeed, our grading maps provide quantitative value reflecting the disease severity, while explanation maps give qualitative information about the relative importance of each feature during the decision-making process. For example, in an explanation map of an AD subject, we do not know if the non-highlighted regions (structures unused by the models to take their decision) are healthy or just non-informative (redundant information with other structures, too noisy due to high inter-subject variability, etc.). Moreover, the explanation maps are generally normalized to the same range of values [0, 1], making the comparison of two explanation maps only qualitative.

This paper is among a few studies proposing an interpretable model for AD classification problem. Another approach for an interpretable model is to carefully design the graph neural network classifier and its input. Li \ea define individual graphs using features extracted from neuroimaging data and a graph neural network with ROI-aware convolutional layer and an appropriate loss function \cite{li_braingnn_2020}. Similar to our result, this approach can provide both salient brain regions at the subject level and the community level. However, similar to explainable methods discussed above, it cannot provide quantitative information on the disease severity for a given region.

While the performance of our framework across unseen datasets and unknown tasks was quite high, there also exist some limitations. First, the ground truth used for grading was potentially not optimal due to a lack of consensus on structures relevant to Alzheimer's disease. Indeed, there may be some structures that are not impacted by AD and the ground-truth of these structures should be zero. With our ground-truth annotation, small structures surrounding another one highly related to AD had a high chance to appear together in all patches. Thus, those structures would be also predicted as related to AD. Another direction should focus on an unsupervised learning manner to find only abnormalities caused by AD to improve the interpretability. Second, this study exploited only structural MRI while the performance could be improved using multi-modal inputs such as PET, functional MRI, diffusion MRI or perfusion MRI \cite{bron_multiparametric_2017}. Better disease patterns are expected to be learned with this kind of input. However, a multi-modal input implies even larger differences between different datasets. Thus, a new generalization study should be considered to see if the gain in performance from better disease patterns can overcome the performance drop resulting from differences between different datasets.

\section{Conclusion}
In this paper, we addressed three major limitations of CNN-based methods by introducing a novel interpretable, generalizable and accurate deep grading framework. First, deep grading offers a meaningful visualization of the disease progression. Second, we proposed a collective artificial intelligence strategy to improve the generalization on datasets owning differences such as MRI protocols, age ranges, country of origin and inclusion criteria. Finally, we proposed to use a graph-based modeling to better capture AD signature using both inter-subject similarity and intra-subject variability. Based on that, our DG method showed state-of-the-art performance in both AD diagnosis and prognosis.

\section*{Acknowledgement}
This work benefited from the support of the project DeepvolBrain of the French National Research Agency (ANR-18-CE45-0013). This study was achieved within the context of the Laboratory of Excellence TRAIL ANR-10-LABX-57 for the BigDataBrain project. Moreover, we thank the Investments for the future Program IdEx Bordeaux (ANR-10-IDEX-03-02), the French Ministry of Education and Research, and the CNRS for DeepMultiBrain project.

Datasets ADNI1-2 used for this project was funded by the Alzheimer's Disease Neuroimaging Initiative (ADNI) (National Institutes of Health Grant U01 AG024904) and DOD ADNI (Department of Defense award number W81XWH-12-2-0012). ADNI is funded by the National Institute on Aging, the National Institute of Biomedical Imaging and Bioengineering, and through generous contributions from the following: AbbVie, Alzheimer’s Association; Alzheimer’s Drug Discovery Foundation; Araclon Biotech; BioClinica, Inc.; Biogen; Bristol-Myers Squibb Company; CereSpir, Inc.; Cogstate; Eisai Inc.; Elan Pharmaceuticals, Inc.; Eli Lilly and Company; EuroImmun; F. Hoffmann-La Roche Ltd and its affiliated company Genentech, Inc.; Fujirebio; GE Healthcare; IXICO Ltd.; Janssen Alzheimer Immunotherapy Research \& Development, LLC.; Johnson \& Johnson Pharmaceutical Research \& Development LLC.; Lumosity; Lundbeck; Merck \& Co., Inc.; Meso Scale Diagnostics, LLC.; NeuroRx Research; Neurotrack Technologies; Novartis Pharmaceuticals Corporation; Pfizer Inc.; Piramal Imaging; Servier; Takeda Pharmaceutical Company; and Transition Therapeutics. The Canadian Institutes of Health Research is providing funds to support ADNI clinical sites in Canada. Private sector contributions are facilitated by the Foundation for the National Institutes of Health (www.fnih.org). The grantee organization is the Northern California Institute for Research and Education, and the study is coordinated by the Alzheimer’s Therapeutic Research Institute at the University of Southern California. ADNI data are disseminated by the Laboratory for Neuro Imaging at the University of Southern California. The OASIS Cross-Sectional project (Principal Investigators: D. Marcus, R, Buckner, J, Csernansky J. Morris) was supported by the following grants: P50 AG05681, P01 AG03991, P01 AG026276, R01 AG021910, P20 MH071616, and U24 RR021382. The Australian Imaging, Biomarker \& Lifestyle Flagship Study of Ageing (AIBL) was supported by the Science \& Industry Endowment Fund. The MIRIAD dataset is made available through the support of the UK Alzheimer’s Society (Grant RF116). The original data collection was funded through an unrestricted educational grant from GlaxoSmithKline (Grant 6GKC).

\bibliographystyle{bibstyles/model1-num-names}
\bibliography{cas-refs}

\pagebreak
\pagebreak
\appendixpage

\subsection*{Performance measures using the AUC metric}
This section presents the various performance measures of the paper using the AUC metric.

\begin{small}
\LTcapwidth=\textwidth
\setlength{\tabcolsep}{4pt}
\begin{longtable}{cc*{6}{>{\centering\arraybackslash}p{1.2cm}}>{\centering\arraybackslash}p{1.4cm}>{\centering\arraybackslash}p{1.2cm}}
\caption{Comparison of different types of features for classification. All the edges are set to 1, the classifier used is GCN. {\color{red} Red}: best result, {\color{blue} Blue}: second best result. The Area Under the ROC Curve (AUC) is used to assess the model performance. The results are the average accuracy of 10 repetitions and presented in percentage. All the methods were trained on the AD/CN subjects of the ADNI1 dataset. Value in bold: $p$ of one-sided Wilcoxon test comparing with baseline (in gray) is lower than 0.05, meaning a significantly superior performance is found compared to the baseline.}\label{table5_auc}\\
\toprule
\multirow{3}{*}{No.} & \multirow{3}{*}{Features} & \multicolumn{4}{c}{\makecell{Diagnosis \\ (AD/CN)}} & \multicolumn{2}{c}{\makecell{Prognosis \\ (p/sMCI)}} & \multicolumn{1}{c}{\makecell{Global \\ Diagnosis \\ (AD/CN)}} & \multicolumn{1}{c}{\makecell{Global \\ Prognosis \\ (p/sMCI)}} \\
& & ADNI2 & AIBL & OASIS & MIRIAD & ADNI1 & AIBL & All & All\\
& & $N=330$ & $N=279$ & $N=756$ & $N=69$ & $N=300$ & $N=32$ & $N=1434$ & $N=332$\\

\midrule

1 & $DG_I$ & {\color{red} \textbf{97.3}} & 94.8 & 93.1 & 99.5 & 74.8 & 74.6 & 95.6 & 74.3\\
2 &  $DG_{Cnw}$ & {\color{blue} 96.8} & 96.5 & {\color{red} \textbf{95.3}} & {\color{red} 100.0} & {\color{blue} 76.7} & {\color{blue} 77.1} & {\color{red} \textbf{96.2}} & {\color{blue}76.6}\\
3 & $DG_C$ & 96.5 & 96.4 & {\color{red} \textbf{95.3}} & {\color{red} 100.0} & 76.6 & {\color{blue} 77.1} & {\color{red} \textbf{96.2}} & {\color{blue}76.6}\\
4 & $V$    & 71.2 & 75.7 & 79.7 & 78.3 & 58.1 & 62.1 & 76.3 & 58.7\\
5 & $A$    & 54.2 & 55.1 & 38.3 & 48.8 & 49.9 & 44.7 & 44.7 & 49.5\\
6 & $V, A$ & 68.0 & 68.5 & 57.7 & 64.9 & 53.6 & 56.4 & 60.9 & 53.8\\
7 & $DG_C, V$ & 95.8 & {\color{blue} 96.9} & {\color{blue} \textbf{94.5}} & {\color{blue} 99.9} & 76.5 & {\color{blue} \textbf{77.6}} & {\color{blue} 95.7} & 76.5\\
\rowcolor{Gray}
8 & $DG_C, A$ & 96.6 & {\color{red} 97.5} & 93.3 & {\color{red} 100.0} & {\color{red} 77.4} & 76.5 & 95.2 & {\color{red} 77.0}\\
9 & $DG_C, V, A$ & 95.9 & {\color{red} 97.5} & 92.8 & {\color{blue} 99.9} & {\color{red} 77.4} & 76.9 & 94.8 & {\color{red} 77.0}\\

\bottomrule
\end{longtable}
\end{small}

\begin{small}
\LTcapwidth=\textwidth
\setlength{\tabcolsep}{4pt}
\begin{longtable}{cc*{6}{>{\centering\arraybackslash}p{1.2cm}}>{\centering\arraybackslash}p{1.4cm}>{\centering\arraybackslash}p{1.2cm}}
\caption{Comparison of different graph edge types. The classifier used is GCN and the input features is $DG_C$ and A. {\color{red} Red}: best result, {\color{blue} Blue}: second best result. The Area Under the ROC Curve (AUC) is used to assess the model performance. The results are the average accuracy of 10 repetitions and presented in percentage. All the methods were trained on the AD/CN subjects of the ADNI1 dataset.}\\ 
\begin{tabular*}{0.97\textwidth}{@{\extracolsep{\fill}}ccccccccc}
\toprule
\multirow{3}{*}{Edge} & \multicolumn{4}{c}{\makecell{Diagnosis \\ (AD/CN)}} & \multicolumn{2}{c}{\makecell{Prognosis \\ (p/sMCI)}} & \makecell{Global \\ Diagnosis \\ (AD/CN)} & \makecell{Global \\ Prognosis \\ (p/sMCI)} \\
& ADNI2 & AIBL & OASIS & MIRIAD & ADNI1 & AIBL & All & All\\
& $N=330$ & $N=279$ & $N=756$ & $N=69$ & $N=300$ & $N=32$ & $N=1434$ & $N=332$\\

\midrule
Fully-one & {\color{blue} 96.6} & {\color{red} 97.5} & {\color{blue} 93.3} & {\color{red} 100.0} & {\color{red} 77.4} & 76.5 & {\color{red} 95.2} & {\color{red} 77.0}\\
Correlation & {\color{red} 96.8} & {\color{blue} 97.4} & 93.0 & {\color{red} 100.0} & {\color{blue} 77.3} & {\color{red} 76.9} & 94.1 & 76.8\\
Volume difference & {\color{red} 96.8} & {\color{red} 97.5} & {\color{red} 93.4} & {\color{red} 100.0} & {\color{blue} 77.3} & {\color{blue} 76.6} & {\color{blue} 94.4} & {\color{blue} 76.9}\\

\bottomrule
\end{tabular*}
\end{longtable}
\end{small}

\pagebreak
\begin{small}
\LTcapwidth=\textwidth
\setlength{\tabcolsep}{4pt}
\begin{longtable}{cc*{6}{>{\centering\arraybackslash}p{1.2cm}}>{\centering\arraybackslash}p{1.4cm}>{\centering\arraybackslash}p{1.2cm}}
\caption{Comparison of different classifiers. For graph-based approaches (\ie all the approaches except SVM and multi-layer perceptron), the edge based on structure volume difference is used and the input features is $DG_C$ and A. {\color{red} Red}: best result, {\color{blue} Blue}: second best result. The Area Under the ROC Curve (AUC) is used to assess the model performance. The results are the average accuracy of 10 repetitions and presented in percentage. All the methods were trained on the AD/CN subjects of the ADNI1 dataset.}\label{table:architecture_comparison_auc}\\
\begin{tabular*}{0.97\textwidth}{@{\extracolsep{\fill}}ccccccccc}
\toprule
\multirow{3}{*}{Classifier} & \multicolumn{4}{c}{\makecell{Diagnosis \\ (AD/CN)}} & \multicolumn{2}{c}{\makecell{Prognosis \\ (p/sMCI)}} & \makecell{Global \\ Diagnosis \\ (AD/CN)} & \makecell{Global \\ Prognosis \\ (p/sMCI)} \\
& ADNI2 & AIBL & OASIS & MIRIAD & ADNI1 & AIBL & All & All\\
& $N=330$ & $N=279$ & $N=756$ & $N=69$ & $N=300$ & $N=32$ & $N=1434$ & $N=332$\\

\midrule
SVM & 94.9 & 95.5 & {\color{red} 93.8} & {\color{blue} 99.9} & 76.1 & {\color{red} 77.1} & 93.7 & 76.1\\
Multi-layer perceptron & 90.4 & 92.8 & 91.0 & {\color{blue} 99.9} & 73.0 & 74.9 & 89.7 & 72.8\\
Transformer & 96.4 & 96.6 & {\color{blue} 93.6} & {\color{blue} 99.9} & 77.1 & 75.3 & {\color{red} 94.6} & 76.6\\
SAGE & {\color{blue} 96.7} & {\color{blue} 97.4} & 93.0 & {\color{blue} 99.9} & 77.1 & 76.0 & 94.1 & 76.6\\
ResGatedGraph & 84.6 & 87.6 & 81.9 & 92.7 & 70.5 & 71.0 & 84.0 & 70.4\\
GAT & 96.6 & 97.2 & 92.7 & {\color{red} 100.0} & {\color{red} 77.5} & {\color{blue} 76.9} & 93.6 & {\color{red} 77.0}\\
TAG & 96.6 & 97.0 & 92.9 & {\color{blue} 99.9} & 77.1 & 76.8 & 94.0 & 76.7\\
GCN & {\color{red} 96.8} & {\color{red} 97.5} & 93.4 & {\color{red} 100.0} & {\color{blue} 77.3} & 76.6 & {\color{blue} 94.4} & {\color{blue} 76.9}\\

\bottomrule
\end{tabular*}
\end{longtable}
\end{small}

\begin{small}
\LTcapwidth=\textwidth
\setlength{\tabcolsep}{6pt}
\begin{longtable}{l*{6}{>{\centering\arraybackslash}p{1.2cm}}}
\caption{Comparison of our method with state-of-the-art methods that have been retrained on our training dataset using the available code and tested on our dataset. {\color{red} Red}: best result, {\color{blue} Blue}: second best result. The Area Under the ROC Curve (AUC) is used to assess the model performance. All the methods are trained on the AD/CN subject of the ADNI1 dataset, the same training/testing partition is used for evaluation.}\label{table6_auc}\\
\toprule
\multirow{2}{*}{Method} & \multicolumn{4}{c}{\makecell{Diagnosis \\ (AD/CN)}} & \multicolumn{2}{c}{\makecell{Prognosis \\ (p/sMCI)}}\\
& \makecell{ADNI2 \\ $N=330$} & \makecell{AIBL \\ $N=279$} & \makecell{OASIS \\ $N=756$}& \makecell{MIRIAD \\ $N=69$}& \makecell{ADNI1 \\ $N=300$} & \makecell{AIBL \\ $N=32$}\\

\midrule

Patch-based CNN \cite{wen_convolutional_2020} & 79.3 & 86.8 & 87.8 & 88.6 & 65.5 & 52.5\\
ROI-based CNN \cite{wen_convolutional_2020} & 90.8 & 90.8 & 92.7 & 97.4 & 69.6 & {\color{blue} 75.0}\\
Subject-based CNN \cite{wen_convolutional_2020} & 85.4 & 90.4 & 92.4 & 98.8 & 70.0 & 59.6\\
Voxel-based SVM \cite{wen_convolutional_2020} & {\color{blue} 93.8} & {\color{blue} 93.6} & {\color{red} 93.6} & {\color{blue} 99.4} & {\color{blue} 74.3} & {\color{blue} 75.0}\\

\midrule
Our method & {\color{red} 96.6} & {\color{red} 97.5} & {\color{blue} 93.3} & {\color{red} 100.0} & {\color{red} 77.4} & {\color{red} 76.5}\\

\bottomrule
\end{longtable}
\end{small}

\begin{small}
\LTcapwidth=\textwidth
\setlength{\tabcolsep}{6pt}
\begin{longtable}{l*{6}{>{\centering\arraybackslash}p{1.2cm}}}
\caption{Comparison of our method with state-of-the-art methods using published results. {\color{red} Red}: best result, {\color{blue} Blue}: second best result. The Area Under the ROC Curve (AUC) is used to assess the model performance. All the methods are trained on the AD/CN subject of the ADNI1 dataset. However, there are many different factors: number of subjects in training/testing sets, selection criteria, etc.}\label{table7_auc}\\
\toprule
\multirow{2}{*}{Method} & \multicolumn{4}{c}{\makecell{Diagnosis \\ (AD/CN)}} & \multicolumn{2}{c}{\makecell{Prognosis \\ (p/sMCI)}}\\
& ADNI2 & AIBL & OASIS & MIRIAD & ADNI1 & AIBL \\

\midrule

Landmark-based CNN \cite{liu_landmark-based_2018} & {\color{blue} 95.9} & - & - & {\color{blue} 97.2} & - & -\\
Hierarchical FCN \cite{lian_hierarchical_2020} & 95.1 & - & - & - & {\color{red} 78.1} & -\\
$AD^2A$ \cite{albarqouni_attention-guided_2020} & 93.4 & {\color{blue} 92.5} & - & - & - & -\\
Efficient3D \cite{yee_construction_2021} & - & - & - & - & - & -\\

\midrule
Our method & {\color{red} 96.6} & {\color{red} 97.5} & {\color{red} 93.3} & {\color{red} 100.0} & {\color{blue} 77.4} & {\color{red} 76.5}\\

\bottomrule
\end{longtable}
\end{small}

\pagebreak

\subsection*{Cross-brain regions connectivity analysis}
\begin{figure}[h]
        \centering
        \includegraphics[width=\linewidth,trim={5cm 1.5cm 5cm 1.5cm}, clip]{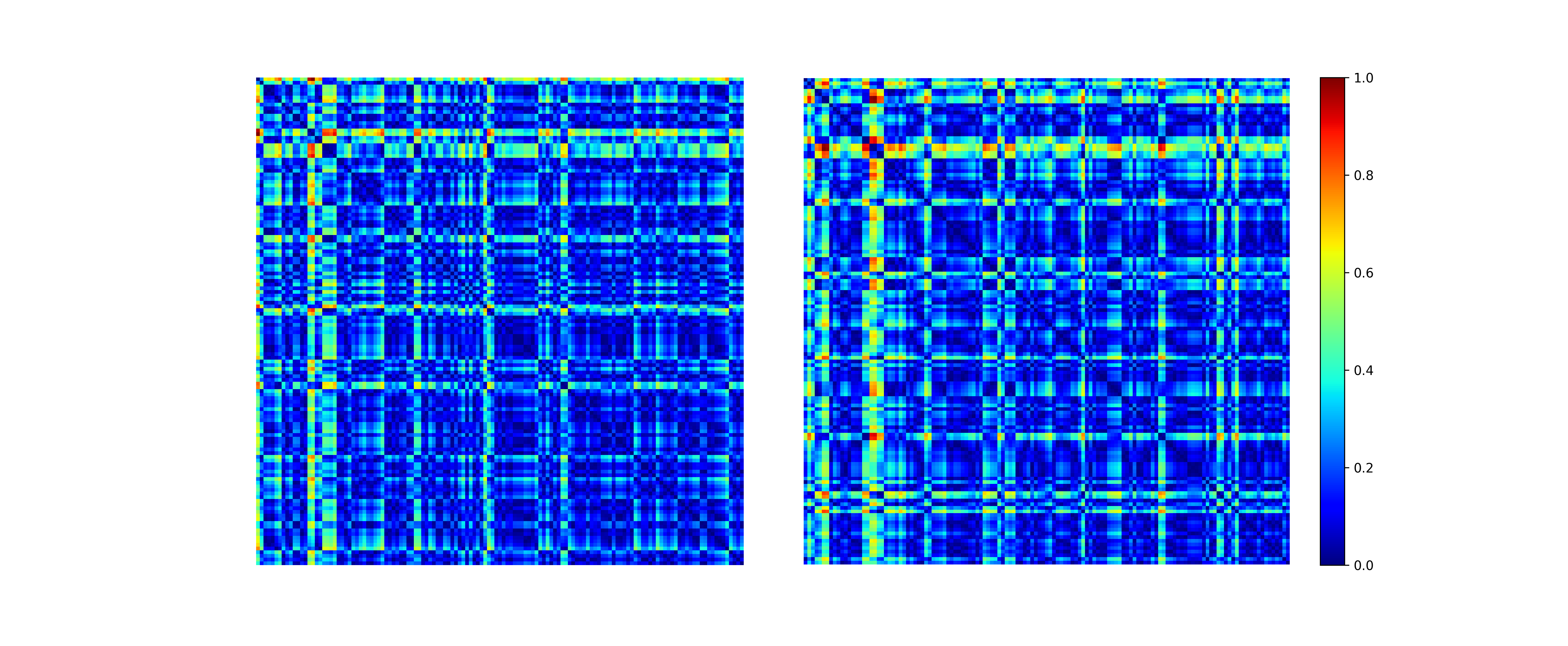}
        \caption{Averaged adjacency matrices of AD population (left) and CN population (right). All the values are normalized to [0, 1] for visualization.}
        \label{figure:adjacency_matrices}
\end{figure}

\noindent
To analyze the cross-brain regions connectivity, we compute two averaged adjacency matrices (\ie edge weights) respectively for all AD patients and all CN subjects using the absolute difference of volumes. After that, we compute the absolute difference of these two matrices (see Figure \ref{figure:adjacency_matrices}). This results in a matrix of size 133 $\times$ 133, we then select 25 highest values (top 0.14\% highest values). These values correspond to 25 pairs of structures. Among these pairs of structures, we observe some structures that have been presented in Section \ref{section:interpretation_dg_maps}, such as bilateral hippocampus, left amygdala, left parahippocampal gyrus and left ventral diencephalon. These structures have been shown to be related to AD \cite{coupe_lifespan_2019, frisoni_clinical_2010, kesslak_quantification_1991, addneuromed_consortium_education_2012}. In AD patients, these structures may present more atrophy volumes than other structures. In CN people, the atrophy volumes of these structures (due to the normal aging process) may be close to other structures. Thus, the absolute difference volumes should be a discriminative feature for AD classification. And in our case, using the absolute difference volumes as the edge weights allows an improvement in performance.

\end{document}